\documentclass[aps,pra,twocolumn,superscriptaddress,nofootinbib,showpacs,floatfix]{revtex4-1}

\usepackage{times}
\usepackage{bm}
\usepackage{graphicx}
\usepackage{color}

\DeclareMathAlphabet{\mathscr}{OT1}{pzc}{m}{it}

\newcommand{\expect}[1]{{\left\langle #1 \right\rangle}}

\newcommand{\dens}[1]{{\rm [#1]}}

\newcommand{\tHe}{\mbox{$^3$He}}

\newcommand{\pinf}{P_{\infty}}
\newcommand{\pspin}{{\sf h}}

\def \ket#1{{\left|#1\right\rangle}}

\def \expect#1{{\left \langle #1 \right\rangle}}

\newcommand{\be}{\begin{eqnarray}}
\newcommand{\ee}[1]{\label{#1}\end{eqnarray}}
\def\favg#1{{\widetilde{#1}}}

\def\hel{{\sf h}}

\begin{document}

 \title{Circular Dichroism of RbHe and RbN$_2$ Molecules}

\def\Wisc{Department of Physics, University of Wisconsin-Madison, Madison, WI 53706, USA}
\def\Julich{Juelich Centre for Neutron Science, Garching 85747, Germany
}
\author{B. Lancor}\affiliation{\Wisc}
\author{E. Babcock}\affiliation{\Julich}
\author{R. Wyllie}\affiliation{\Wisc}
\author{T. G. Walker}\affiliation{\Wisc}

\date{\today}

\begin{abstract}We present measurements of the circular dichroism of optically pumped Rb vapor near the D1 resonance line.  Collisions with   the buffer gases $^3$He and N$_2$  reduce the transparency of the vapor, even when fully polarized.  We use two methods to measure this effect, show that the He results can be understood from RbHe potential curves, and show how this effect conspires with the spectral profile of the optical pumping light to increase the laser power demands for optical pumping of very optically thick samples.
\end{abstract}
\maketitle

Optical pumping of alkali-metal atoms at high temperatures and high buffer gas pressures is a powerful technique for precision spectroscopies (clocks, magnetometers) \cite{Jau04,Kominis03}, and for collisional transfer of angular momentum to noble gas nuclei via spin-exchange collisions \cite{WalkerRMP}.  The ability to spin-polarize large quantities of nuclei using spin-exchange optical pumping (SEOP) normally requires the alkali-metal vapors to be $\sim$100 optical depths at line center.  The light is only able to penetrate such dense samples if the atoms cease to absorb light once they are fully spin-polarized \cite{Bhaskar79}.  It is for this reason that optical pumping at high optical thickness is only possible using circularly polarized light resonant with S$_{1/2}\rightarrow$ P$_{1/2}$ (colloquially, ``D1") transitions, where angular momentum conservation  forbids $m_S=1/2$ atoms to absorb the  helicity  $\hel=1$ pumping light.\footnote{Throughout the paper we shall assume the quantization axis is parallel to the pumping light propagation direction.  Thus $\hel=1$ light corresponds to photons with $\expect{J_z}=1.$} In contrast, optical pumping with S$_{1/2}\rightarrow$ P$_{3/2}$ (``D2") light tends to partially polarize the atoms in strongly absorbing states \cite{Happer87}.

This paper presents quantitative measurements of the normally forbidden absorption of $\sf h=1$ D1 light by $m_S=1/2$ Rb atoms in atmospheric pressure cells containing He and N$_2$ buffer gases.  We are led to this investigation by the repeated observations in our lab and elsewhere that the amount of light required for SEOP is substantially greater than models predict  \cite{Chen07}.  Furthermore, previous studies of SEOP with Rb-K mixtures observed that alkali-metal polarizations saturated  significantly below 100\% even for high optical pumping rates  \cite{Babcock03}.  This suggested a light-intensity dependent spin-relaxation rate.  Finally, when pumping pure Rb with broadband light sources, it was observed that the Rb polarization saturated well below 100\%, again suggesting that the pumping light does not go dark for polarized atoms \cite{Babcock03}.

Since these experiments are typically done with gas pressures of  several atmospheres, it is natural to ask to what extent pressure broadening gives a small amount of D2 character to the D1 resonance. Several previous studies suggest this. First, spectroscopic measurements show that the red wing of the D2 line certainly overlaps the D1 resonance  \cite{Drummond74}.
In addition,  He collisions can collisionally transfer 5P$_{1/2}$ population to 5P$_{3/2}$, albeit with a small cross section  \cite{Rotondaro98}.  The existence of this excited-state spin-relaxation process implies that during collisions with He atoms the
5P$_{1/2}$ and 5P$_{3/2}$ states are somewhat mixed.
Finally, from a theoretical perspective it is well known that the fine-structure interaction is partially decoupled by the intramolecular fields, thereby mixing P$_{1/2}$ and P$_{3/2}$ molecular states \cite{Allard82}.

\begin{figure}[h]
\includegraphics[width=3.0 in]{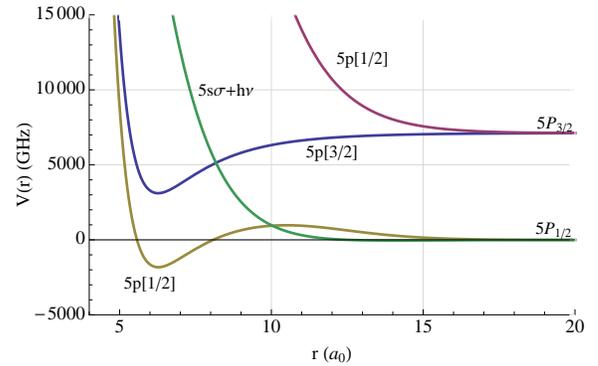}
\caption{Adiabatic energy curves for RbHe molecules adapted from Ref.~\protect \cite{Pascale83} with the method of Ref.~\protect \cite{Allard82}.  The projection of the total electronic angular momentum along the interatomic axes is given in brackets.  The curve crossings between the photon-dressed 5s$\sigma$ state and the two excited-state curves means that photon absorption is allowed during a collision.  As explained in the text, these absorption processes are not subject to the free-atom dipole selection rules, allowing normally angular momentum forbidden transitions to occur. }
\label{fig:pots}
\end{figure}

The partially D2-like absorption of light during RbHe collisions can be understood by considering the lowest adiabatic potential energy curves for Rb-He molecules, shown in Fig.~\ref{fig:pots}.  It is well known  \cite{Allard82} that RbHe collisions are  non-adiabatic, but for a qualitative understanding the adiabatic curves should be useful. The curves that correlate with the two 5p fine structure levels are designated 5p$[\Omega]$, where $\Omega=1/2,3/2$ is the magnitude of the projection of the electronic angular momentum on the interatomic axis.  The $\Omega=3/2$ state has pure P$_{3/2}$ character, while the $\Omega=1/2$ states contain  significant mixtures of the two fine-structure components for interatomic separations $r< 15$ $a_0$.  Also shown in Fig. \ref{fig:pots} is the ground state potential $5s\sigma$ shifted up by the energy of a D1 photon.  The crossing between the 5p$[3/2]$ state and the repulsive wall of the photon-shifted $5s\sigma$ state at {$r=8$ $a_0$} means that resonant  D2-like light absorption can occur during  RbHe collisions.  Since $kT/h=9400$ GHz at typical 180$^\circ$C temperatures for SEOP, the curve crossing at an energy of $h\times$ 5000 GHz is readily thermally accessed.  It is also seen from Fig. \ref{fig:pots} that the fine-structure-mixed 5p[1/2] molecular potential is resonantly accessed by D1 light.

In the following we present circular dichroism measurements of Rb atoms broadened by pure N$_2$ buffer gas and He rich He-N$_2$  buffer gas mixtures (typical of SEOP cells), near the Rb first resonance lines.  When the dichroism measurements are combined with known pressure-broadened lineshapes, the results show that the light absorption cross section for $\hel=1$ D1 light by fully polarized atoms is $\sigma^{1}=1.49\pm0.15\times 10^{-17}$ cm$^2$ at a N$_2$ density of 1 amg  \cite{amg}, and  $\sigma^{1}=1.10\pm0.12\times 10^{-17}$ cm$^2$ at a He density of 1 amg.  Measurements at two He pressures confirm the RbHe molecules as the source of the absorption.  Despite the small size of these cross sections in comparison to on-resonant cross sections of $1\times 10^{-12}$ cm$^2$, they are sufficient to significantly lower the alkali-metal polarization and increase the power required to optically pump optically thick vapors, especially for broad-band light sources.

This paper, a full account of Ref.~\cite{Lancor10},  is organized as follows.  In Section~\ref{sec:narrow} we discuss the circular dichroism of spin-1/2 atoms, and discuss its consequences for optical pumping of optically thick vapors.  In Section~\ref{sec:expt} we present our measurements of circular dichroism, using two quite different methods.  We then present in Section~\ref{sec:theory} a theoretical estimate of the effect and show that the expected size  is in accord with the experimental results.  Section~\ref{sec:model} is a more detailed discussion of the consequences  of the measurements for optical pumping.  Section~\ref{sec:hfs}  discusses complications from hyperfine interactions that appear for a small portion of the experimental data.  

\section{Effects of Reduced Circular Dichroism on Optical Pumping of Thick Vapors}\label{sec:narrow}

To facilitate some of the discussions later in the paper, we include here a brief discussion of the relationship between circular dichroism and optical pumping.  We will assume monochromatic light for this section.  A more complete analysis, including key spectral averaging effects, will be given in Sec.~\ref{sec:model}.

Circular dichroism is the differential absorption of light of opposite helicities.  To avoid the difficult problem of determining absolute atom densities, it is convenient to measure a normalized circular dichroism, defined as
\be
{\cal C}={\expect{\sigma^{-1}}-\expect{\sigma^{1}}\over \expect{\sigma^{-1}}+\expect{\sigma^{1}}}
\ee{cd}
where the average is over the various populations $\rho_m$ of the magnetic sublevels of the atom,
\be
\expect{\sigma^\pspin}=\sum_m\rho_m\sigma^\pspin_m,
\ee{sigdef}
and $\sigma_m^\pspin$ is the absorption cross section for light of helicity $\pspin=\pm1$ by an atom in sublevel $m$.
The atoms are assumed to be purely longitudinally polarized parallel to a bias magnetic field.

The general form of the light absorption cross section for  spin-1/2 atoms  is \cite{Babcock03}
\be
{\sigma^{\pspin P}}=\expect{\sigma^\pspin}=\sigma_0(1-P_\infty \pspin P )
\ee{gensig} where the atomic polarization is $P=\rho_{1/2}-\rho_{-1/2}$ and the light is assumed to propagate parallel to the magnetic field.  The absorption cross section for unpolarized light is $\sigma_0$. 
For now, we neglect hyperfine structure. Since the absorption cross section depends only on the product of the helicity and the spin-polarization, the dichroism can be measured by reversing the direction of  either the helicity or the polarization. The parameter  $\pinf$  is limited to the range $-1\le\pinf\le1$. Using Eq.~\ref{gensig} in Eq.~\ref{cd}, it follows that
\be
{\cal C}=P\pinf.
\ee{normcirc}
Thus $\pinf$ is the normalized circular dichroism of a fully spin-polarized atom.  For  alkali-metal atoms $P_\infty(\nu_1)\approx 1$ for light near the center of the D1 line, and $P_\infty(\nu_2)\approx -1/2$ at the D2 line\cite{Happer87}.  Therefore ${\cal C}\approx P$ for an alkali-metal vapor probed by resonant D1 light. We shall refer to the condition ${\cal C}=P$, or equivalently $P_\infty=1$, as ``full dichroism".

The value of $\pinf$ also affects the optical pumping process.  Under conditions of large pressure broadening, full electron spin-randomization in the excited state, and nuclear spin conservation in the excited state, the optical pumping process for monochromatic $\pspin=1 $ light obeys \cite{Babcock03}

\be
{d\expect{F_z}\over dt}={R\over 2}(\pinf- P)-{\Gamma P\over 2}
\ee{optpump}
where $\expect{F_z}$ is the total angular momentum of the atom (electronic plus nuclear), $R=\sigma_0I/h\nu$ is the optical pumping rate from light of intensity $I$, and $\Gamma$ is the ground-state spin-relaxation rate which may be due to a variety of collision processes.  Thus we see that the steady-state polarization is
\be
P=\pinf {R\over \Gamma+R}
\ee{steadystatepol}
so that the maximum polarization that is attainable at infinite pumping rate is $P=\pinf$.

The steady-state absorption rate is
\be
A&=&R(1-\pinf P)={\Gamma R\over \Gamma+R}\left[1+{R\over\Gamma}\left(1-{\pinf^2}\right)\right]\\
&\approx&{\Gamma R\over \Gamma+R}\left[1+{2R\over\Gamma}(1-\pinf)\right]
\ee{absrate}
where the approximation holds for $1-\pinf\ll1$. For full  dichroism, the atoms absorb at the rate $\Gamma P$, the amount required to repolarize the atoms due to ground-state spin-relaxation.  For reduced dichroism, the scattering rate increases by the factor
\be
\Upsilon\approx \left[1+{2R\over\Gamma}(1-\pinf)\right]
\ee{upsilon}
which can be much greater than 1 when the spin-polarization is high ($R\gg\Gamma$).
This condition is necessary for pumping of optically thick vapors, where the rule of thumb is that in order to fully polarize the cell  volume, the pumping rate at the cell entrance must be at least the optical depth times the relaxation rate.

\begin{figure}
\includegraphics[width=3.0 in]{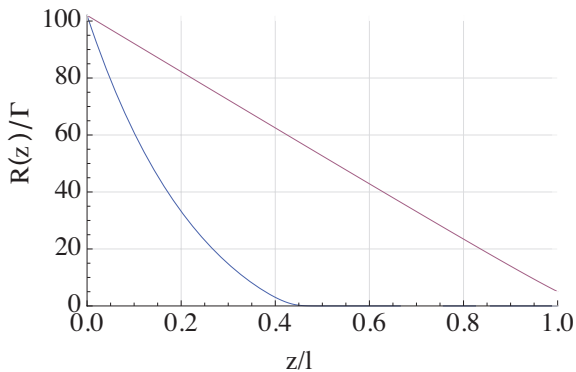}
\caption{Optical pumping rate as a function of propagation distance $z$, for a monochromatic pumping beam.  In the case of full dichroism, the light is attenuated only by the need to repolarize atoms that undergo spin-relaxation collisions.  When polarized atoms continue to absorb light due to imperfect dichroism, the light attenuation is much more rapid.  The cell length $l$ was chosen so that the vapor is 100 optical depths thick.  }
\label{fig:lightprop0}
\end{figure}

To illustrate the effects of reduced dichroism, Fig.~\ref{fig:lightprop0} compares monochromatic light propagation for full dichroism to that with a modest $\pinf=0.98$, for an optical thickness of 100.  For the ideal  case, the intensity steadily decreases at the rate governed by spin-relaxation.  The reduced dichroism case sees an initial intensity decrease that is five times larger, resulting in the light being able to polarize only about 40\% of the atoms.

\section{Dichroism Measurements}\label{sec:expt}

In this section we present the two methods used to measure the normalized circular dichroism of Rb--buffer gas molecules.

\subsection{Direct Optical Method}

We deduced the normalized  circular dichroism of fully polarized atoms $\pinf$ by comparing the absorption of positive helicity light by atoms with polarizations $\pm P$ using the apparatus and methods described here.

\begin{figure}
\includegraphics[width=3.0 in]{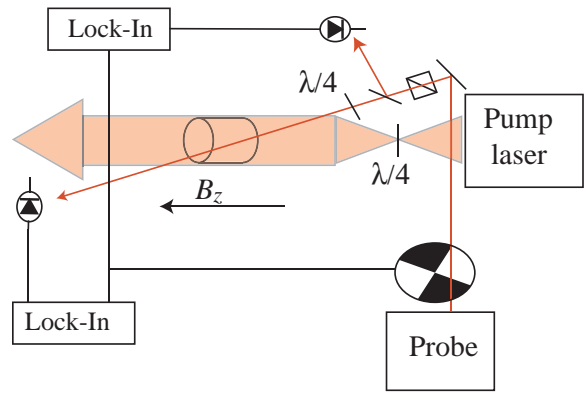}
\caption{Apparatus for measuring the circular dichroism of Rb-He vapor.Ê The pump laser, propagating parallel to a magnetic field, spin-polarizes Rb atoms by optical pumping.  The polarization is set either parallel or antiparallel to the field, as determined by the orientation of a quarter-wave plate.Ê The fractional transmission of a weak, circularly polarized, tunable probe laser is determined by the ratio of photodiode voltages before and after traversal of the cell.Ê The circular dichroism is then determined from the transmissions for both directions of Rb polarization.Ê The absolute Rb polarization is measured with the apparatus of Fig.~\ref{fig:eprexp}.}
\label{fig:exp}
\end{figure}

A schematic of the apparatus is shown in ÊFig.~\ref{fig:exp}. ÊÊA Rb vapor cell, contained in a flowing hot-air oven, was optically pumped by a circularly polarized frequency narrowed diode array bar  providing 35 Watts of power at 795nm, with a spectral width of of $\sim$125 GHz \cite{Babcock05}. ÊA holding field of 50 G was applied in the pump propagation direction. ÊA probe beam, one of two single-frequency external cavity diode lasers  tunable around 780 nm and 795 nm, was attenuated to P$< 50$ $\mu$W, sent through a chopper operating at 485 Hz, and linearly polarized with a polarizing beam splitter cube. ÊDirectly in front of the oven, the beam went through a non-polarizing beam splitter plate to provide a reference signal proportional to the incident intensity, and a quarter wave plate to produce circular polarization. ÊThe reference and transmitted intensities were measured on silicon photo-diodes and sent to lock-in amplifiers referenced to the chopper frequency. ÊTo change the direction of the atomic spin polarization relative to the probe helicity, the pump ${\lambda/4}$ plate was rotated 90$^\circ$, thus reversing the pump laser helicity. ÊTo obtain the relation between the incident and transmitted intensities Êin the absence of Rb (thus accounting for loss in the windows of the oven and cell), a measurement was taken at room temperature. 

Three natural abundance Rb cells were used in this experiment. ÊThe pure N$_2$ cell is a 4.5 cm diameter blown Pyrex sphere, with 2.80 amg N$_2$  \cite{amg}, the low pressure SEOP cell is a closed 4.9 cm long cylinder (Corning 1720 body with GE180 windows), containing 0.80 amg $^3$He and 0.07 amg  of N$_2$. The high pressure SEOP cell is a blown GE180 sphere of diameter 3.5 cm, Êfilled with 3.27 amg of $^3$He and 0.13 amg N$_2$.  ÊThe temperature of the cell was monitored by an infrared thermocouple, and controlled by an analog controller. ÊTemperatures ranging from $\sim$ 60 $^\circ$C to $\sim$ 180 $^\circ$C, corresponding to [Rb]= $1-200\times 10^{12}$ cm$^{-3}$, were used to produce appropriate optical thickness for transmission measurements at a range of frequencies. At the lower temperatures, the pump laser power was reduced to lower background noise.  
 
For the pure N$_2$ cell, Rb liquid  droplets on the walls of the cell produced an uncontrolled temperature dependence and drift to the probe transmission as the droplets slowly moved on the face of the cell.   To account for this, the  probe laser and an additional 856 nm  laser were coupled into a fiber, with the 856 nm laser transmission serving to measure small variations of the transmission of the cell walls with time.

\begin{figure}
\includegraphics[width=3.0 in]{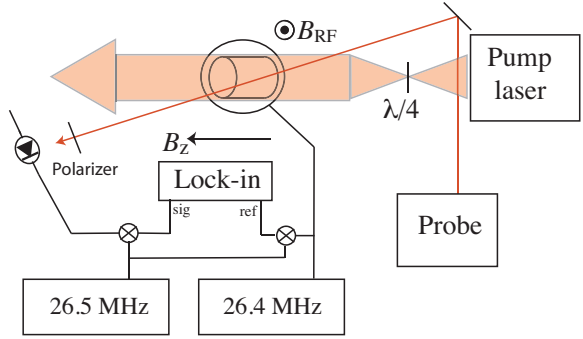}
\caption{Measurement of polarization via transverse EPR spectroscopy.  As the DC magnetic field is swept to match the Rb Zeeman EPR frequencies, the index of refraction of the Rb atoms is modulated at the 26.4 MHz RF driving frequency, thus modulating the polarization of the probe laser.  This modulation is detected and demodulated by the lock-in amplifier.  The Rb polarization is deduced from the ratio of areas of the EPR resonances.}
\label{fig:eprexp}
\end{figure}

The Rb spin polarization Êwas measured with transverse electron paramagnetic resonance (EPR) spectroscopy \cite{Baranga98,Appelt98}, as shown in Fig.~\ref{fig:eprexp}. A 26.4 MHz RF field was applied perpendicular to the holding field by driving a pair of 9 cm diameter coils, separated by 7 cm, with a Êsynthesized function generator. ÊThe amplitude was controlled with a voltage controlled attenuator. ÊAs the holding field was swept through the EPR resonances of either isotope, the transverse component of the probe beam, now linearly polarized, acquired a polarization modulation at the RF frequency, which was converted to an intensity modulation with a Êpolarizer oriented at 45$^\circ$ to the original probe polarization direction. ÊThe probe laser frequency was detuned from the D2 resonance  1-2 nm so that it had negligible effect on the optical pumping. The probe intensity was measured by a fast photodetector, the output of which was amplified by two RF amplifiers. ÊThis signal was demodulated to 100 KHz by mixing it with a 26.5 MHz signal from a second synthesized function generator. ÊThis 100 KHz signal was then sent to a lock-in amplifier referenced to a signal generated by mixing the outputs of the two function generators. ÊThe phase of the lock-in was chosen to produce Lorentzian EPR spectra, which were recorded in the lock-in and sent to a computer for analysis. ÊThe spin-polarization was deduced using the area ratio method of Ref.~ \cite{Baranga98}. ÊThis method is insensitive to the broadening mechanism of the EPR lines, and in particular can be applied for both low and high spin-exchange rates, {\it i.e.} over a wide range of temperatures.  All the polarization measurements were done with $^{85}$Rb,  with corresponding measurements made on $^{87}$Rb at low temperatures where the two species were not in spin-temperature equilibrium.

The probe beam propagates through the cell at an angle $\theta=17.59\pm0.2^\circ$. ÊThe modulated component of the Faraday rotation is proportional to $P_x\sin\theta$ and is therefore much smaller than would be detected by a transverse probe. ÊHowever, we wish to know the polarization averaged along the probe beam direction, so there was a tradeoff between the EPR signal, proportional to $\sin\theta$ Êand the DC circular dichroism, which is degraded by a factor $\cos\theta$. ÊWe chose $\theta$ to be as small as possible and still observe high quality EPR signals. ÊAt high densities, this is not a problem, but for line-center measurements where the temperatures (as low as 70 $^\circ$C) are not sufficient to produce spin-temperature equilibrium  \cite{Happer72}, it was essential to observe many EPR peaks in order to back out the polarization.

We wish to find the normalized circular dichroism, from which we can deduce $\pinf$. ÊThe basic parameters observed by the experiment are the transmissions $I_\pm$ for  circularly polarized light of helicity $\pspin$ propagating at angle $\theta$ to the atomic polarization $\pm P$. ÊThen the absorption cross section for the probe light becomes \cite{Chann02d}
\be
\sigma^{\pspin P}(\theta)=\sigma_0(1-P_\infty \pspin P \cos\theta)
\ee{sigtheta}
Since the absorption cross section  is proportional to $\pspin P$,
reversing the direction of the atomic spins is equivalent to reversing the helicity of the probe. ÊWe found it advantageous to reverse the pump polarization and leave the probe helicity fixed, as the light transmission of the oven and cell windows depended slightly on the probe helicity. When the pump polarization was reversed, the atomic polarization satisfied $(1-|P'_{Rb}|)=(1-P_{Rb})$ to within 5\% at worst, and  typically to within much less.  The imperfect atomic polarization reversal was measured and corrected for in the data. We also corrected Êeach intensity measurement to account for an observed residual transmission of the probe beam even at very high optical depths ($\sim 0.5\%$). This probably results from a small amount of light at the free-running probe laser wavelength.
Data points with transmission $< \sim 2\%$ were not used due to the uncertainty in this correction.
A correction was also made for typical probe helicity $\pspin=0.997$.

We Êextract $\pinf$ from the corrected transmitted intensities $I_\pm=I_o\exp(-\dens{Rb}\sigma^{\pm P}(\theta)l)$ by finding
\be
{\cal C}=\frac{-\ln\left(\frac{I_{-}}{I_o}\right)+\ln\left(\frac{I_{+}}{I_o}\right)}{-\ln\left(\frac{I_{-}}{I_o}\right)-\ln\left(\frac{I_{+}}{I_o}\right)}=P P_\infty \cos\theta.
\ee{asymmetry}
Note that in forming this ratio the optical thickness $\dens{Rb}l$ and instrumental gains cancel.
Combining with measurements of $\theta$ and $P$ gives
\be
\pinf={{\cal C}\over P\cos\theta}.
\ee{BLd}
For regions of the data where $\pinf\approx 1$, it is particularly crucial to make an accurate measurement of $1-P$. ÊThis was complicated by the fact that we did not achieve high atomic polarizations at low [Rb], where the on resonance data must be taken. ÊAt [Rb]$\sim 1.5\times 10^{12} $ cm$^{-3}$ we observed $P$ from 0.9 to 0.95 in all three cells, while at [Rb]$> 50\times 10^{12} $ cm$^{-3}$ we observed $P\geq 0.99$ in the He-N$_2$ cells, and $P\sim 0.95$ in the N$_2$ cell. The typical uncertainty of the EPR measurements was $\sim 10\%$ or less in $(1-P)$.

For the near resonance data taken in the spherical cells, the  uncertainty in $P$ was the largest uncertainty.  Etalon interference effects in the walls of the cylindrical cell (0.75 $\%$ to 1.0 $\%$) were comparable to the  uncertainty in $P$ near resonance, and were the dominant contribution to the errors off resonance.   Etalon effects were smaller in the spherical cells ($<0.25 \%$), where drifts in $I_o$ (0.25 $\%$ to {0.5 \%}) dominated off resonance uncertainties for the spherical SEOP cell, and relative intensity drift between the 795 nm and 856 nm lasers (0.5$\%$) was the primary source of off resonant error in the pure N$_2$ cell.  ÊFor most of the frequency range covered, the data was  taken at high density where $P$ was close to 1 and the uncertainty in $P$ was a minor component of the total uncertainty. The uncertainties in the probe propagation angle and the measured probe intensities were small under all conditions.

An important check on the reliability of the measurements was a detailed study of the dichroism very near the D1 line center, where $\sigma^{-1}\gg\sigma^{1}$ giving $\pinf\approx 1$. ÊIn this region, hyperfine interactions are  quite important, especially for the low pressure cell, and a complex structure is observed. ÊThis study will be presented in Sec.~\ref{sec:hfs}. ÊThe results are quite sensitive to the accuracy of the measurement of $(1-P)$, and the successful understanding of the experiment in this region gives additional confidence in the off-resonant results.

\begin{figure}
\includegraphics[width=3.5 in]{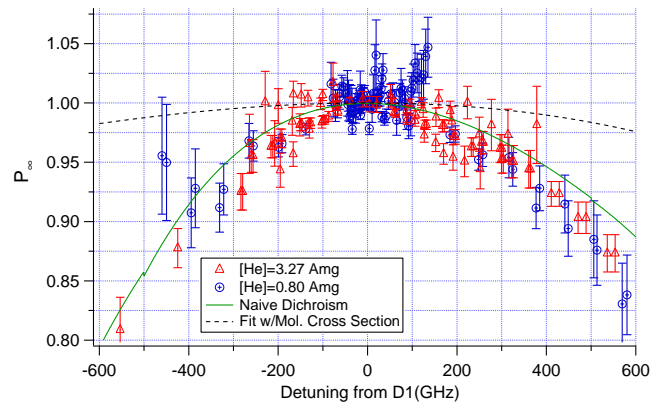}
\caption{Normalized circular dichroism results near the D1 resonance line. ÊThe agreement between cells of different He pressure verifies that the effects originates from absorption in RbHe collisions. ÊThe dashed  line denotes the frequency dependence of the dichroism making the naive assumption of purely Lorentzian line broadening, and the solid line is the result of using the molecular absorption cross section from Eq.~\ref{fitfun}.}
\label{fig:res2}
\end{figure}

The measured  $\pinf$ near resonance, from transmission data, for both SEOP cells is shown in ÊFig.~\ref{fig:res2}. ÊThe circular dichroism deviates significantly from one within the typical 1000 GHz  bandwidth of broadband sources such as diode array bars often used for SEOP. ÊThe implications of this finding will be discussed in Sec.~\ref{sec:model}. ÊA very important result from Fig.~\ref{fig:res2} is the Êagreement between the two cells, despite their very different He pressures. ÊAt detunings outside the atomic linewidth of 15-60 GHz  \cite{Romalis97}, the absorption of negative helicity light is proportional to the buffer gas pressure. ÊThus, only if the normally forbidden absorption of positive helicity light is also proportional to the buffer gas pressure will the dichroism be pressure independent. ÊThe agreement between the two cells at different pressures  confirms that the source of the impure dichroism in these cells is  Rb--buffer-gas collisions.

\begin{figure}
\includegraphics[width=3.5 in]{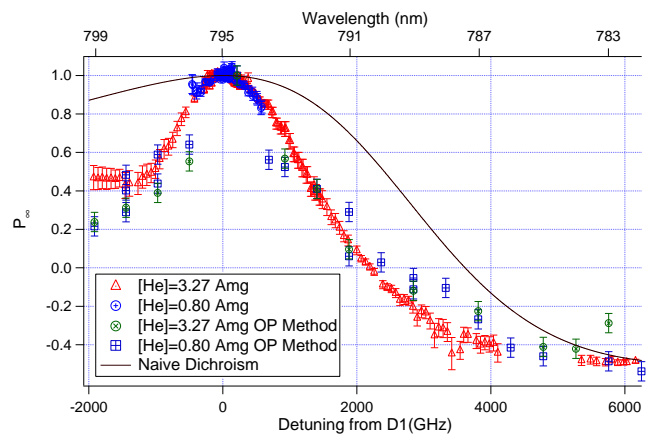}
\caption{Measured circular dichroism of RbHe molecules in the region between the Êfirst resonance lines of Rb. ÊThe solid curve shows the expected dichroism from the very naive assumption of purely Lorentzian broadened lines. }
\label{fig:wide2}
\end{figure}

For the high \dens{$^3$He}  cell, we were able to measure $P_\infty$ across the entire frequency range from the D1 to the ÊD2 resonance, and this is plotted Êin Fig.~\ref{fig:wide2}. (The gap in the data from 784-6 nm is due to lack of coverage of that range by our two probe lasers.) ÊAs expected, the dichroism goes from $+1$ to $-0.5$ as the light is tuned between the two resonances. ÊHowever, the sign of the dichroism flips at $790.325\pm 0.03$ nm,  a significant shift from the $787.5$ nm zero crossing that would result from a naive  model that takes the cross-section at each frequency to be the sum of  Lorentzian pressure-broadened D1 ($\pinf=1$) and D2 ($\pinf=-1/2$)  lineshapes. Ê

\begin{figure}
\includegraphics[width=3.5 in]{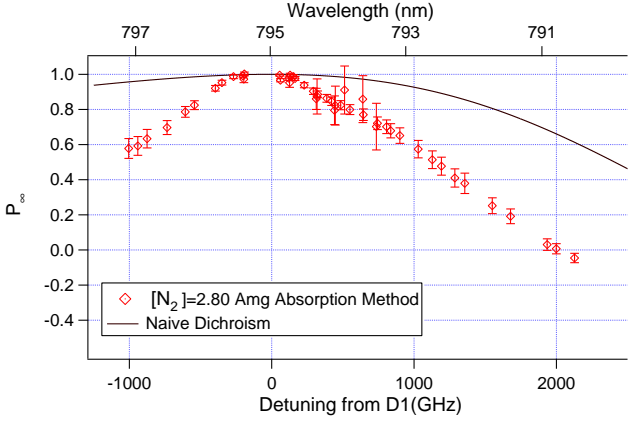}
\caption{Measured circular dichroism of RbN$_2$ molecules in the region between the Êfirst resonance lines of Rb. ÊThe solid curve shows the expected dichroism from the very naive assumption of purely Lorentzian broadened lines. Ê}
\label{fig:N2Dic}
\end{figure}

To allow us to isolate the contributions of Rb-N$_2$ collisions, we measured $P_\infty$ in the pure N$_2$ cell (Fig. ~\ref{fig:N2Dic}).  Pure He cells cannot be used for these measurements because some N$_2$ is necessary  to eliminate radiation trapping \cite{WalkerRMP}, so these pure N$_2$ results allow us to subract out the  small  N$_2$ contribution in the SEOP cells.

\begin{figure}
\includegraphics[width=3.5 in]{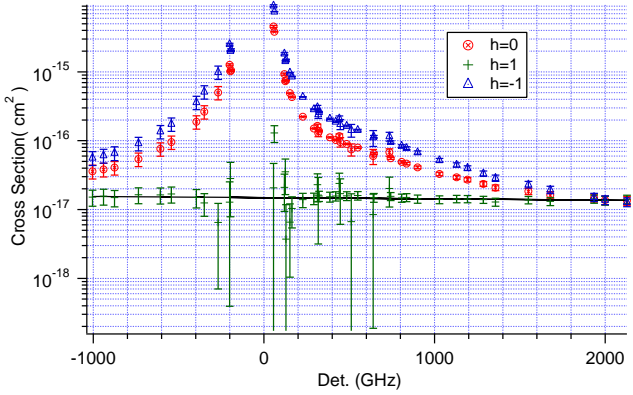}
\caption{Cross sections, normalized to 1 amg N$_2$ density, for absorption of $\sigma^1,\sigma^{-1}, \text{and }\sigma^0$ light by fully polarized Rb atoms. The solid line is the linear fit to the $\sigma^1$ data.}
\label{fig:Ncross}
\end{figure}

Given the importance of these measurements for D1 optical pumping, it is convenient to extract the Êabsorption cross section for positive helicity light by fully polarized atoms. ÊTo do this, we derive the product
\be
\dens{Rb}l\sigma^0=\frac{-\ln\left({I_+/I_o}\right)-\ln\left({I_-/I_o}\right)}{2}
\ee{BLi}
from the transmission data, and fit it near resonance to the expected lineshape from  \cite{Romalis97} to find $\dens{Rb}l$, $l$ being the probe propagation distance through the cell. ÊOur measured lineshapes agree well with Ref.~ \cite{Romalis97} in the near-resonant region.  The N$_2$ data deviates from this lineshape at detunings more than $\sim$ 100 GHz. ÊGiven $\sigma^0$ and $\pinf$, we then deduce $\sigma^{\pm 1}$ using Eq.~\ref{gensig}. ÊThe three cross-sections for Rb-N$_2$ are shown in Fig.~\ref{fig:Ncross}. 

\begin{figure}
\includegraphics[width=3.5 in]{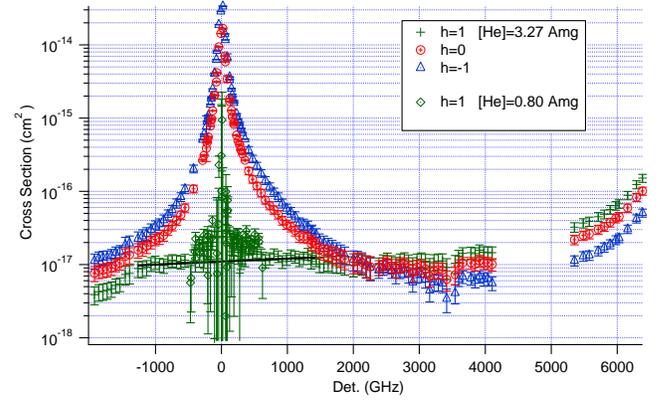}
\caption{Cross sections, normalized to 1 amg He density, for absorption of $\sigma^1,\sigma^{-1}, \text{and }\sigma^0$ light by fully polarized Rb atoms. The solid line is a linear fit to $\sigma^1$ in the vicinity of the D1 resonance.}
\label{fig:cross2}
\end{figure}

Assuming the N$_2$ cross-section is proportional to \dens{N$_2$}, we fit  $\sigma^1$ to a line in the region near the D1 line, giving
\be\sigma^1(\nu)&=&\sigma^1(\nu_1)+(\nu-\nu_1)\left.{d\sigma^1\over d\nu}\right|_{\nu_1} \label{fitfun}\\
{\sigma^1(\nu_1)\over \dens{N_2}}&=&1.49\pm 0.15\times10^{-17} {\mbox{cm}^2\over \mbox{amg}}, \\
{1\over \dens{N_2}}\left.{d\sigma^1\over d\nu}\right|_{\nu_1}&=&-6\pm 5\times10^{-22} {{\rm cm}^2\over {\rm amg}\,\,\rm{ GHz}}.
\ee{results}
We  then use this result to extract the normalized cross section at the D1 line for He-Rb from the SEOP
 cells, finding
 \be
{\sigma^1(\nu_1)\over \dens{He}}&=&1.10\pm 0.12\times10^{-17} {\mbox{cm}^2\over \mbox{amg}},\\
{1\over \dens{He}}\left.{d\sigma^1\over d\nu}\right|_{\nu_1}&=&6\pm 5\times10^{-22} {{\rm cm}^2\over{\rm amg}\,\,\rm{ GHz}}.
\ee{results2}
The Rb-N$_2$ cross sections from Fig.~\ref{fig:Ncross} were also used along with the data from the SEOP cells to get the Rb-He cross sections across the full frequency range.  These are shown in
Fig.~\ref{fig:cross2}.

The $\sigma^0$ measurements in Fig.~\ref{fig:cross2} are to our knowledge the first far-wing absorption measurements of the RbHe molecule. Previously, Gallagher and coworkers  \cite{Drummond74,Ottinger75} measured near- and far-wing emission spectra, and pointed out the non-quasi-static nature of the RbHe emission spectrum.  If we assume that the D2 red-wing emission spectrum roughly corresponds to the contribution to $\sigma^1$ from the $5p[3/2]$ state,
we would deduce $3\times 10^{-18}$ cm$^2$ from Ref.~ \cite{Drummond74}.  As will be discussed in Sec.~\ref{sec:theory}, calculations suggest that the  $5p[3/2]$ contributes about 1/3 of $\sigma^1$.  This is in excellent agreement with our measurements, though this should be considered somewhat fortuitous given that a number of assumptions have been made in comparing the two experiments.

\subsection{Optical Pumping Method}

As a second, quite different, method for measuring $\pinf$ at frequencies off the D1 resonance, we used a 30 W frequency narrowed (50 GHz linewidth) external cavity laser  \cite{Babcock05} to optically pump the atoms at different frequencies $\nu$ and measured the resulting polarization $P(\nu)$, pumping rate $R(\nu)$, and spin-relaxation rate $\Gamma$. This method was used only on the SEOP cells. For off-resonant pumping, we assume $\pinf$ is essentially constant for the small range of frequencies present in the pumping beam, so Eq.~\ref{steadystatepol} becomes
\be
\pinf(\nu)=\left(1+{\Gamma\over R(\nu)}\right)P(\nu).
\ee{EarlPinf}
For most of the data, $\Gamma\gg R(\nu)$.

\begin{figure}[htb]
\includegraphics[width=3.0in]{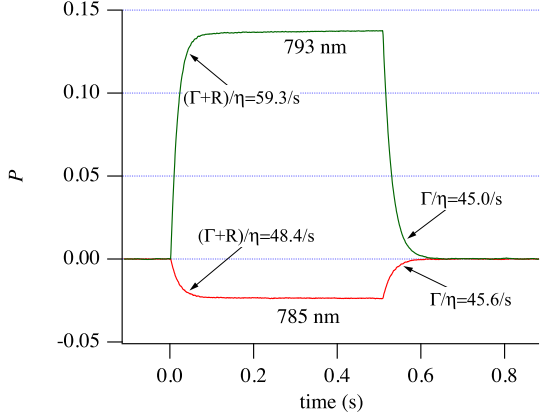}
\caption{Pumping/decay transients used to measure $\pinf$. Pumping light is turned on at $t=0$, and the polarization builds up to the steady-state value of Eq.~\ref{EarlPinf}.  The exponential build-up and decay constants allow the pumping rate $R$ and the relaxation rate $\Gamma$ to be measured.  Note that the sign of the polarization reverses for 785 nm pumping as compared to 793 nm pumping, due to the dichroism of the vapor being negative for the former case. }
\label{fig:EarlTransient}
\end{figure}

We chopped the pumping laser with a mechanical shutter and measured the spin-polarization as a function of time, as illustrated in Fig.~\ref{fig:EarlTransient}, using Faraday rotation \cite{Happer72,Chann02c}.  For small polarizations, the rising transient  builds up polarization to the steady state value (\ref{EarlPinf}) with a rate constant $(R(\nu)+\Gamma)/\eta$, and the falling transient decays with  a rate constant $\Gamma/\eta$, where the slowing-down factor $\eta=10.8$ for natural abundance Rb takes into account the spin inertia due to the alkali-metal nuclei at low polarization in spin-temperature equilibrium  \cite{Appelt98}.  The Faraday rotation was calibrated by pumping on the D1 resonance, where longitudinal EPR spectroscopy \cite{Chann02c} verified that the polarization was very close to 100\%.

We took data using two separate  frequency narrowed diode array bars, one with a free
running wavelength of 794 nm at room temperature and the other with a free running
wavelength of 800 nm at room temperature. With these two bars, narrowed  and frequency shifted by
our external  cavity design \cite{Babcock05}, we were able to access wavelengths from 782 nm to 810 nm by also changing the temperature
 from
-20$^\circ$C to 45$^\circ$C using a  water-cooled thermo-electric mount.

The measured values of $\pinf$ using this optical pumping method are shown in Fig.~\ref{fig:wide2}.  They agree quite well with the direct circular dichroism measurements described above.  The  two SEOP cells were used for the measurements, and again we find that the dichroism is the same within errors for the two cells, consistent with the RbHe molecular absorption hypothesis.

With this method,  the zero crossing of the dichroism is particularly dramatic as the signal of Fig.~\ref{fig:EarlTransient}
reverses sign near 790 nm.

\section{Estimates}\label{sec:theory}

In this section we will present estimates of the dichroism of RbHe molecules in the vicinity of the D1 resonance line.  To our knowledge, the circular dichroism of these molecules has never been studied before, nor have calculations been published.  A full calculation is beyond the scope of this work, but the considerations given here will attempt to explain our observations.

We consider the light absorption process as a collision, where the ground-state potential curve is shifted up  by one photon energy.  As seen from Fig.~\ref{fig:pots}, 795 nm light can be resonantly absorbed to the 5p[3/2] state during a collision.  In the absence of fine-structure mixing by the atom-atom interactions, the 5p[1/2] state that correlates at large $r$ to the P$_{1/2}$ state would not absorb the $\sigma^1$ light from a spin-polarized Rb atom.  However, the fine-structure interaction causes the 5p[1/2] states to be mixtures of P$_{1/2}$ and P$_{3/2}$ character, with the P$_{3/2}$ part giving rise to an allowed light absorption.  In this case, the Rabi coupling will depend on  the interatomic separation $r$, going to zero at large $r$.

To simplify the calculations, we will assume that we can neglect the variation in the light-atom coupling with collision angle, replacing the angle-dependent Rabi coupling with its angular average.  We will also assume that we can neglect rotational coupling of the adiabatic curves  \cite{Allard82}.  With these assumptions, the light-absorption process is isotropic and we can consider the problem as a simple collision in the radial coordinate, with excitation to each of the excited-state potential curves considered separately.

Assuming a classical trajectory with distance of closest approach $r_0$ (assumed to occur at time $t=0$),  it is convenient to define
\be
\alpha(r)&=&\int_{r_0}^r {dr\over v(r)}\Delta V(r)\\
v(r)&=&v_\infty\sqrt{1-{b^2\over r^2}-{2V_{5s\sigma}(r)\over mv_\infty^2}}
\ee{alpha}
where $K=mv_\infty^2/2$ is the initial kinetic energy, corresponding to initial speed $v_\infty$, and $b$ is the impact parameter for the collision.  The difference potential is $\hbar\Delta (r)=V_e(r)-V_{5s\sigma}(r)-\hbar\omega$.  Then, in the weak intensity limit, first order perturbation theory gives the probability of finding the atom in the excited state $e$ after a single  collision is
\be
|c_e|^2&=&\left|\int_{r_0}^\infty{dr\over v}\epsilon(r)\cos\alpha(r)\right|^2.
\ee{ce}
The molecular Rabi frequency is given in terms of the atomic Rabi frequency as
\be
\epsilon(r)={\epsilon_0\over 2}\sin(\beta(r)/2)
\ee{rabi}
where the factor of 2 accounts for the angular average over collision orientations, and $\beta$ is the P$_{3/2}$ mixing angle to be discussed below.
From this we get the excitation rate to a single adiabatic potential by averaging over the collision impact parameters:
\be
{\cal R}=\dens{He}v_\infty\int \pi  db^2|c_e|^2=\sigma^1I/\hbar\omega
\ee{exrate}
 The atomic Rabi frequency $\epsilon_0$ is related to the light intensity and the atomic lifetime $\tau$ by $\epsilon_0^2=3\lambda^3I/(2\pi h c \tau)$, giving finally
\be
\sigma^1(v_\infty)={3\lambda^2v_\infty\over 2\pi\tau}\dens{He}\int \pi db^2{|c_e|^2\over \epsilon_0^2}.
\ee{sig1avg}
This expression is then averaged over a Maxwellian velocity distribution.

We begin by considering absorption to the 5p[3/2] curve in Fig.~\ref{fig:pots}.  This state is of pure P$_{3/2}$ character, so that $\beta=\pi$.  D1 light is sufficiently far detuned that non-adiabatic effects should not be too important, and we consider the process as a Landau-Zener transition from the ground state 5s$\sigma$+$h\nu$ at the crossing point.  The probability of excitation is
\be
P_{LZ}={\pi\hbar\epsilon^2\over v_r |d\Delta/dr|},
\ee{LZ}
with  all quantities evaluated at the crossing point.  Performing the impact parameter integration, the thermal average, and accounting for the two-fold $|\Omega|$ degeneracy  then gives
\be
\sigma_{\rm opt}&=&\dens{He}{3\pi^{3/2}\lambda^2 r_c^2\over 8\tau|d\Delta/dr|}e^{-V_{5s\sigma}/T}.
\ee{LZresult}
For 1 amg of He, this evaluates to $1.8\times 10^{-18}$ cm$^2$ at the D1 resonance,  about a factor of 2 smaller than a direct numerical integration of Eq.~\ref{ce} that gives $3.9\times10^{-18}$ cm$^2$.

\begin{figure}[htb]
\includegraphics[width=2.5 in]{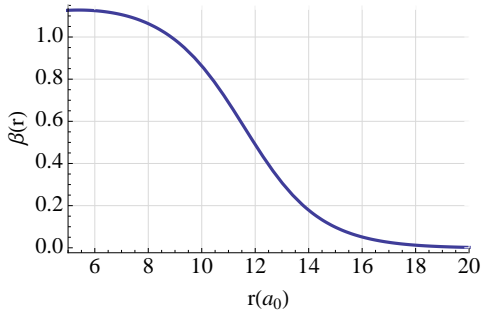}
\caption{Fine-structure mixing angle for the 5p[1/2] states as a function of interatomic separation.}\label{fig:phase}\end{figure}

For the 5p[1/2] curve that correlates to the P$_{1/2}$ state at $r=\infty$, the Landau-Zener approximation is invalid because the phase $\alpha$ does not vary rapidly enough with $r$.  The $r-$dependent wavefunction, written in terms of the atomic fine-structure states $\ket{JM}$, is $\ket{5p[1/2]}=\cos(\beta/2)\ket{{1\over 2}{1\over 2}}+\sin(\beta/2)\ket{{3\over 2}{1\over 2}}$.  The calculated mixing angle $\beta(r)$ is shown in Fig.~\ref{fig:phase}.  Numerical integration of Eqs.~\ref{ce} and \ref{sig1avg} gives $5.5\times 10^{-18}$ cm$^2$ at 1 amg.

Thus our estimate of the absorption cross section $\sigma^1=9.4\times 10^{-18}$ cm$^2$ at 1 amg is very close to the measured value.  Given that we have ignored rotational coupling of the adiabatic curves and the anisotropic nature of the light-molecule Rabi coupling, this excellent agreement may be considered somewhat fortuitous.  Nevertheless, it confirms that it is quite reasonable for the circular dichroism of RbHe molecules to be at the level we observe in the experiment.

\section{Consequences of Reduced Circular Dichroism for Spin-Exchange Optical Pumping}\label{sec:model}

We now turn to the implications of these results for spin-exchange optical pumping of \tHe.  With the advent of high power diode array bars having 10s to 100s of Watts of low-cost power available at 795 nm, it has become standard practice to use these lasers for SEOP \cite{Driehuys96}.  One drawback of these lasers is that they have relatively broad spectral widths, typically several nm (1 nm=475 GHz at 795 nm).  This was shown to be mitigated by the practice of running at multi-atmosphere He pressures, where the atomic line could be broadened and directly interact with a more substantial fraction of the laser linewidth \cite{Driehuys96}.  However, the highest \tHe\ polarizations, around 80\%, have only been obtained with spectrally narrowed lasers \cite{Chann00,Chann03,Chen07}.  We shall see that the reduced dichroism effect explains much of this behavior, since the off-resonant light of an un-narrowed laser accentuates the decrease in dichroism.  In addition, the frequency spectrum of the light changes as the light propagates through the cell, as the resonant portions of the spectrum are preferentially attenuated, while the off-resonant portions experience little attenuation.  This causes a further decrease in the effective dichroism as the light propagates.

We begin by generalizing the results of Sec.~\ref{sec:narrow} to pumping sources with non-zero bandwidth.
The absorption rate of $\pspin=1$ light by  spin-1/2 atoms of electron spin-polarization $P$ is
\be
\favg{A}=\favg{R}-\favg{R\pinf} P
\ee{absrate2}
where $\favg{f}=\int d\nu f(\nu)$.  The resulting optical pumping evolution is given by
\be
{d\expect{F_z}\over dt}={1\over 2}\left ({\favg{R\pinf}}- \favg{R}P-{\Gamma P}\right)
\ee{optpump2}
In steady-state, the equilibrium polarization reached by the atoms is
\be
P={\favg{R\pinf}\over \Gamma+\favg{ R}}.
\ee{pol}
We see from Eq.~\ref{pol} that in the limit of large pumping rates, the alkali-metal polarization saturates at the pumping-rate-weighted mean value of $\pinf$.

The steady-state absorption rate is found by substituting Eq.~\ref{pol} in Eq.~\ref{absrate2} to get
\be
\favg{A}&=&{\Gamma\favg{R}\over \Gamma+\favg{R}}\left[1+{\favg{R}\over\Gamma}\left(1-{\left[\favg{R\pinf}\over\favg{R}\right]^2}\right)\right]\\
&\approx&{\Gamma\favg{R}\over \Gamma+\favg{R}}\left[1+{2\over \Gamma}\favg{R\delta_\infty}\right]\;\;(\delta_\infty\ll 1)
\ee{?}
where $\delta_\infty=1-\pinf$ measures the deviation of $\pinf$ from its maximum value of 1.  If $\delta_\infty=0$, the ideal case, the absorption rate becomes  $\Gamma P$, and the light dissipation is simply the amount required to compensate for the collisional losses.  At high pumping rates, as required for optical pumping of optically thick vapors, the factor $2\favg{R\delta_\infty}/\Gamma$ can be large and the light is absorbed by the atoms at a much greater rate.

To explore the consequences of reduced dichroism for optical pumping at high densities, we have simulated the optical pumping and light propagation effects described above.  At each point in the optical pumping cell, the pumping rate is calculated from the spectral profile of the light at that position, and the atomic polarization is calculated from the pumping rate, $\pinf$, and the ground-state spin-relaxation rate (calculated as described in Ref.~ \cite{Chen07}) according to Eq.~\ref{pol}.  Then the spectral profile of the light   is    propagated to the next place in the cell using
\be
{dI(\nu)\over dz}=-\dens{Rb}\sigma(\nu)I(\nu)\left[1-P\pinf(\nu)\right].
\ee{lightprop}
The diffusion layer at the front of the cell is accounted for in an approximate manner using the model of  Ref.~ \cite{WalkerRMP}.  Heating effects  \cite{Walter01} have been neglected.  The electron spin is assumed to completely relax in the excited state, and the nuclear spin is assumed to be conserved in the excited state.

\begin{figure}[htb]
\includegraphics[width=3.0in]{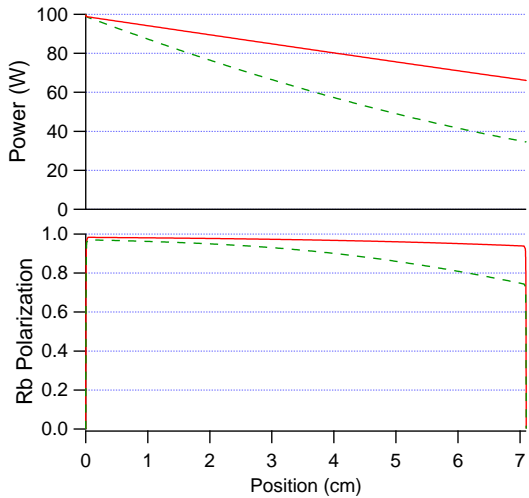}
\caption{Light propagation results with full (solid red) and measured (dashed green) circular dichroism, for pumping with a laser whose spectral profile is shown in Fig.~\ref{fig:spect}.  The upper graph shows the power as a function of position.  The reduced dichroism produces a faster attenuation of the light.  The lower graph shows the corresponding Rb polarizations, which are limited to values less than 100\% even at the cell entrance, but then decay further as the spectral profile of the light is increasingly off-resonant (see Fig.~\ref{fig:spect}).}
\label{fig:broad}
\end{figure}

\begin{figure}[htb]
\includegraphics[width=3.0in]{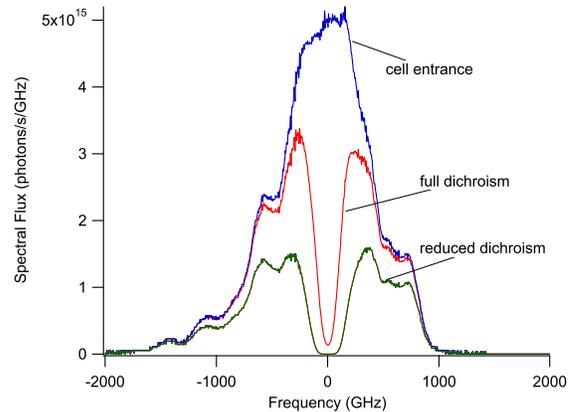}
\caption{Measured spectra before and modeled spectra after propagation through the cell  of Fig~\ref{fig:broad}, modeled as described in the text.  Light out in the wings of the spectrum disproportionately contributes to the reduced dichroism, while contributing comparatively less to the pumping rate.  The net effect is significantly greater absorption than expected.}
\label{fig:spect}
\end{figure}

The effects of reduced dichroism are most readily seen for broad-band pumping light at high \dens{He}, shown in Fig.~\ref{fig:broad}.
We assume 100 W of broadband pumping light (spectral profile shown in Fig.~\ref{fig:spect}) enters a 10 cm diameter, 7 cm long cell with 8 amg of \tHe\ and 50 Torr of N$_2$, with \dens{Rb}=$4\times 10^{14}$ cm$^{-3}$.
Under these conditions we estimate a spin-relaxation rate of 630/s \cite{Chen07}.  In the full dichroism case the light is attenuated only due to ground-state spin-relaxation and the 35 W dissipated in the cell is consistent with this.   The polarization is maintained at a very high level, averaging 97\%, as the pumping rate has not been sufficiently reduced to cause a substantial Rb polarization drop at the back end of the cell.  The on-resonant portion of the laser spectral profile is not yet completely attenuated, as shown in Fig.~\ref{fig:spect}, again consistent with maintenance of a high pumping rate. When the reduced dichroism is taken into account, several changes occur.  The power dissipation per unit length is substantially increased, as seen in Fig.~\ref{fig:broad}, even at the entrance to the cell before the spectral hole is burned.  The total power dissipation is much greater than in the ideal case, now 65 W.  The polarization drop is now quite substantial, reducing to 75\% at the back of the cell.  This is due to two effects:  1) the pumping rate is lower due to the greater power dissipation  and the production of a complete hole in the spectral profile (shown in Fig.~\ref{fig:spect}), and 2) the remaining light is in the spectral region with low dichroism, further reducing the maximum attainable polarization.

\begin{figure}[htb]
\includegraphics[width=3.0in]{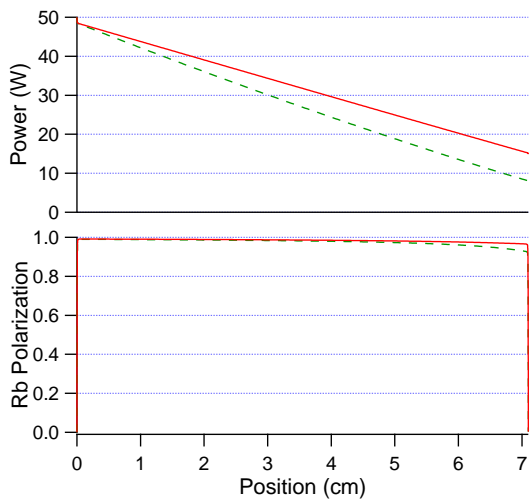}
\caption{Light propagation results with full (solid red) and measured (dashed green) circular dichroism, for a 100 GHz bandwidth pumping source.  The upper graph shows the power as a function of position.  The reduced dichroism produces a slightly faster attenuation of the light, but high polarization is maintained throughout the cell, as shown in the lower graph.}
\label{fig:narrowsim}
\end{figure}

From the above observations, some of the advantages of narrowband pumping are explained, as illustrated in Fig.~\ref{fig:narrowsim} for a 100 GHz laser with half the power of the broadband laser of Figs. \ref{fig:broad} and \ref{fig:spect}.  A source with spectral width of 100 GHz or less will experience  nearly full dichroism, reducing the excess light absorption and increasing the maximum attainable polarization.  Furthermore, as the light is attenuated, the spectral profile does not change significantly and the polarization remains very high. These observations are qualitatively consistent with experimental observations and comparisons of broad and narrow-band sources \cite{Chann03,Chen07}.

\section{Measurements Near the D1 Line Center}\label{sec:hfs}

As an important check on the reliability of the direct optical method, we  studied the dichroism near line center where hyperfine interactions play a role, especially in the low pressure  SEOP cell.  This region is particularly good for testing the accuracy of our polarimetry as our laser, despite its high intensity, was unable to fully spin-polarize the Rb atoms at the low temperatures ($\sim70^\circ$C) needed for moderate optical thickness at the line center.

\begin{figure}[htb]
\includegraphics[width=3.0 in]{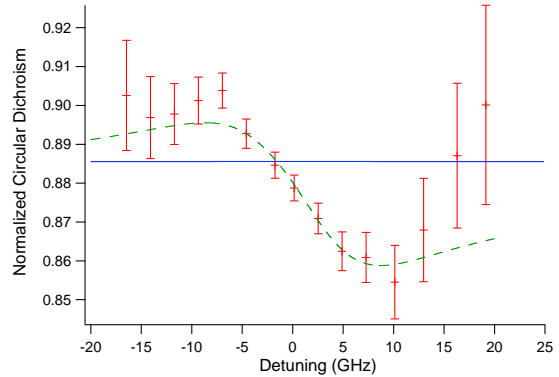}
\caption{Normalized Circular Dichroism near resonance, taken at low density, and low polarization, $P=0.929$.  Modelling that omits the hyperfine (solid blue line) splitting predicts a nearly flat $\cal{C}$ at $P\cos{\theta}$.  Modelling with $\pinf=1$, but including the ground state hyperfine structure (dashed green line), is in good agreement with the data.}
\label{fig:onres2}
\end{figure}

Although the hyperfine splittings (3.0 GHz for $^{85}$Rb, 6.8 GHz for $^{87}$Rb) are smaller than the pressure-broadened atomic  linewidth, the hyperfine effect on the measured normalized circular dichroism is striking, as seen in Fig.~\ref{fig:onres2}.  The low polarization of this data also makes an accurate measurement of $(1-P)$ critical.  Since the atoms are not in spin temperature equilibrium for this region of the data, we found it necessary to be able to resolve EPR peaks that were $<{1/500}$ the size of the primary peak in order to accurately deduce the polarization from the EPR peak areas.  It was also necessary to measure the polarization of the two isotopes independently at low temperatures.

The  shape of $\cal{C}$ near resonance is primarily due to a frequency shift between the absorption cross sections of light of opposite helicity.  An unpolarized vapor has the same cross section for absorption of all helicities of light. In a highly polarized vapor, most of the atoms are pushed into a stretched state in the upper hyperfine manifold, $\ket{F=2,m_F=2}$ for $^{87}$Rb, which absorbs only light of helicity $\pspin=-1$ (for the case of full dichroism).  The rest of the atoms are in other states of high $m_f$ in both the upper and lower hyperfine manifolds.  The high $m_F$  states in the upper manifold are preferentially spin up, while those in the lower manifold are preferentially spin down.  Thus, in a spin-temperature-like distribution, most of the atoms that absorb light of helicity $\pspin=1$ are in the lower manifold, while the overwhelming majority of atoms absorbing light of $\pspin=-1$ are in the upper manifold.  As the two manifolds have slightly different center frequencies, this leads to the structure seen in Fig.~\ref{fig:onres2}.

To model $\cal{C}$ for near resonance conditions, we can write the spin dependence of the cross sections $\sigma^{\pspin}_{Fm_f}$ in terms of the cross sections in the fine structure basis
\be
\sigma^{\pspin}_{m_s}=\sigma_o(1-2 m_s \pinf\pspin)
\ee{sigfs}
Using Clebsch-Gordan coefficients, and neglecting small corrections due to excited-state hyperfine structure,
\be
\sigma^{\pspin}_{Fm_F}(\nu)=\sigma_o(\nu-\nu_F)\left(1\mp{m_F\over F}
\pspin \pinf\right)\\
\ee{sighfs}
for $F=I\pm 1/2$, and the center frequencies of the hyperfine lines are $\nu_F$.  Then the cross sections $\sigma^{\pspin}$ can be built from  the populations $\rho_{Fm_F}$ that are extracted from the EPR data, giving
\be
\sigma^{\pspin}(\nu)=\sum_{Fm_F}\sigma^{\pspin}_{Fm_F}(\nu)\rho_{Fm_F}
\ee{sigFavg}
Modeling that used these cross sections, with $\pinf=1$, is in good agreement with the near resonance data (Fig.~\ref{fig:onres2}).

\section{Conclusions}\label{sec:conclusion}

We have presented here   measurements of the spin-dependence of line broadening of alkali-metal atoms in He and N$_2$ buffer gases. In particular we observe the breakdown of atomic selection rules due to buffer gas collisions \cite{Lancor10}. We presented two very distinct methods for making the measurements that nonetheless gave very similar results.   While line-broadening of alkali-metal atoms by noble gas atoms has been extensively studied  \cite{Allard82}, the spin-dependence has not been previously studied.  We have pointed out in this paper that the unique conditions of spin-exchange optical pumping are quite sensitive to the spin-dependence of far-wing absorption, since such absorption becomes effectively a spin-relaxation mechanism that is proportional to the light intensity.  Such a mechanism implies not only a limit on the maximum polarization that can be obtained, but can greatly increase the laser power requirements.  This effect at least partially accounts for the substantial improvements observed when using frequency narrowed lasers  \cite{Chann03}.

In this paper, we have limited our focus to the effects of reduced dichroism from RbHe and RbN$_2$ collisions.  We have not attempted to compare modeling predictions with SEOP experiments because two other effects, non-conservation of nuclear spin in the excited state and radiation trapping, are also important and have yet to be included in our models.
In upcoming papers, we will discuss these effects and we intend to present an updated experimental comparison to a full model that includes reduced dichroism and these other effects.

\begin{acknowledgments}
Discussions with T. R. Gentile were very helpful. This work was supported by the Department of Energy.
\end{acknowledgments}
\bibliography{/Users/Thad_Walker/Research/thadbibtex/spinexchange}
\end{document}